\documentclass[conference]{IEEEtran}

\pdfoutput=1
\usepackage{amsmath}    
\usepackage{amssymb}
\usepackage{graphicx}   
\usepackage{verbatim}   
\usepackage{color}      
\usepackage{subfig}  
\usepackage[draft]{hyperref}   
\usepackage{mathtools}
\usepackage[normalem]{ulem} 
\newcommand{\stkout}[1]{\ifmmode\text{\sout{\ensuremath{#1}}}\else\sout{#1}\fi}
\usepackage{soul} 
\usepackage{kantlipsum}
\usepackage{comment}
\allowdisplaybreaks

\def \CAP {\textsc{cap}}
\def \ConfigSet {\Gamma}
\def \ConfigIdx {\gamma}
\def \cost {\textsc{cost}}
\def \core {\textsc{core}}
\def \ncore {n^{\textsc{core}}}
\def \ncoref {n^{\textsc{core}}_{f}}
\def \NFV {\textsc{nfv}}

\def \redcost {\textsc{red\_cost}}
\def \SD {\mathcal{SD}}
\def \ServiceChainSet {C}
\def \ServiceChainIdx {c}
\def \zRMP {z_{\ConfigIdx}}
\def \myspace {\hspace*{-1.cm} }

\begin{document}
%
\title{Multiple Service Chain Placement and Routing in a Network-enabled Cloud\\
       \large {\color{green}This is a preprint electronic version of the article 
		submitted to IEEE ICC 2017}}

\author{\IEEEauthorblockN{Abhishek Gupta\IEEEauthorrefmark{1},
Brigitte Jaumard\IEEEauthorrefmark{2}, Massimo Tornatore\IEEEauthorrefmark{1}\IEEEauthorrefmark{3}, and
 Biswanath Mukherjee\IEEEauthorrefmark{1}}
\IEEEauthorblockA{\IEEEauthorrefmark{1}University of California, Davis, USA \ \ 
\IEEEauthorrefmark{2}Concordia University, Canada \ \
\IEEEauthorrefmark{3}Politecnico di Milano, Italy \\
Email: \IEEEauthorrefmark{1}\{abgupta,mtornatore,bmukherjee\}@ucdavis.edu  \IEEEauthorrefmark{2}bjaumard@cse.concordia.ca  \IEEEauthorrefmark{3}massimo.tornatore@polimi.it} }

\maketitle

\begin{abstract}
Network Function Virtualization (NFV) aims to abstract the functionality of traditional proprietary hardware into software as Virtual Network Functions (VNFs), which can run on commercial off-the-shelf (COTS) servers. Besides reducing dependency on proprietary support, NFV helps network operators to deploy multiple services in a agile fashion. Service deployment involves placement and in-sequence routing through VNFs comprising a Service Chain (SC). Our study is the first to focus on the computationally-complex problem of multiple VNF SC placement and routing while considering VNF service chaining explicitly. We propose a novel column-generation model for placing multiple VNF SCs and routing, which reduces the computational complexity of the problem significantly. Our aim here is to determine the ideal NFV Infrastructure (NFVI) for minimizing network-resource consumption. Our results indicate that a Network-enabled Cloud (NeC) results in lower network-resource consumption than a centralized NFVI (e.g., Data Center) while avoiding infeasibility with a distributed NFVI.  
\end{abstract}


%

\section{Introduction}
\label{intro}
Today's communication networks provide services through proprietary hardware appliances (e.g., network functions such as  firewalls, NAT, etc.) which are statically configured to provide a service. Static configurations of network functions are difficult to deploy, modify, and upgrade. Further, with rapid evolution of applications in today's internet, networks require more agile and scalable service deployment. A rapid and flexible deployment of services, however, is not feasible with traditional configured services since they are physically embedded in the network. 

Network Function Virtualization (NFV) \cite{etsi_nfv} provides network operators with a solution. NFV propagates the idea of hardware functionality abstracted into software modules called Virtual Network Functions (VNFs). VNFs can be run on commercial-off-the-shelf (COTS) hardware such as servers and switches. VNF placement on COTS hardware avoids embedding functions in the network and makes service deployment agile and scalable. But VNF placement has resource requirements (e.g., CPU cores, RAM, storage) that need to be satisfied by hardware, same as Virtual Machines (VMs). Since VNFs essentially are VMs, they can be deployed in the cloud.

\begin{figure}
\center
 \includegraphics[scale=.35]{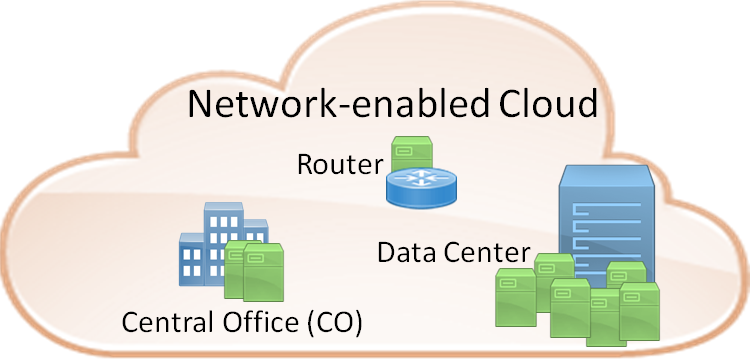}
 \caption{A Network-enabled Cloud (NeC).}
  \label{fig:scNeC}
\end{figure}

Cloud usually refers to massive compute and storage facilities offered by Data Centers (DCs). Here, we extend cloud to the network with a Network-enabled Cloud (NeC) \cite{rt_cloud} where network elements including elements from the underlying optical backbone network that are equipped with computing and storage resources can be part of the NFV Infrastructure (NFVI) and are called ``NFV Point of Presence'' (NFV-PoP) such as router and Central Office (CO) in Fig. \ref{fig:scNeC}.  A NeC will lead to better network resource utilization as traffic flows will take shorter paths by removing frequent redirection to DC for service. Any implementation of NFVI including NeC will rely heavily on the optical backbone network.

When network functions are configured to provide a service, we have a  ``Service Chain''. The term ``service chain'' is used ``to describe the deployment of
such functions, and the network operator's process of specifying an ordered list of service functions that should be applied to a deterministic set of traffic flows'' \cite{ietf_sc}. So, a ``Service Chain'' (SC) specifies a set of network functions configured in a specific order. With NFV, we have VNF SCs where VNFs are configured in a specific sequence to form a SC as shown in Fig. \ref{fig:multiSC}.

In this work, we solve the problem of multiple VNF SC placement and traffic-flow routing. This problem is difficult to scale and computationally complex which is a significant challenge for network operators as they aim to deploy NFV to provide more agile services while utilizing network resources efficiently. We propose a novel sophisticated column-generation model to solve this problem and find the ideal NFVI scheme. The relatively-small computation time of the model helps us to scale to large problem instances which is not possible with standard mathematical modeling. 

The rest of the study is organized as follows. Section \ref{relWrk} gives an overview of major works to solve the VNF placement and routing problem. Section \ref{probDesc} formally describes the problem and its input parameters. We then describe our column-generation model in Section \ref{colGen} to solve the multiple VNF SC placement and routing problem. Section \ref{sim_examples} uses illustrative examples to demonstrate the idea of an NeC. Section \ref{concl} concludes the study.

\section{Related Work}
\label{relWrk}
A number of studies exist on the VNF placement and routing problem. The authors of Ref. \cite{vnf_placement_turck} study a hybrid deployment scenario with hardware middleboxes using an Integer Linear Program (ILP), but do not enforce VNF service chaining  explicitly. Ref. \cite{place_vnf_secci} uses an ILP to study trade-offs between legacy and NFV-based traffic engineering but does not have explicit VNF service chaining. Ref. \cite{vnf_placement_barcellos_gaspary} models the problem in a DC setting using an ILP to reduce the end-to-end delays and minimize resource over-provisioning while providing a heuristic to do the same. Here too the VNF service chaining is not explicitly enforced by the model. Ref. \cite{orc_vnf_boutaba} models the batch deployment of multiple chains using an ILP and develops heuristics to solve larger instances of the problem. However, they enforce that VNF instances of a function need to be on a single machine and restrict all chains to three VNFs. Our model does not impose such constraints, and we allow any VNF type to be placed on any NFV-PoP and any number of VNFs in a SC while service chaining VNFs for a SC explicitly.
 
Our previous work \cite{ilp_report} solved the problem for a single VNF SC in a NeC. Scaling the model to multiple VNF SCs resulted in exponential time complexity. We now address this challenge by using a novel decomposition model (column generation) for multiple VNF SC placement and routing. Our objective is to minimize network-resource consumption in any NFV infrastructure (NFVI) where VNF SCs are deployed. And here, we make the case for a Network-enable Cloud (NeC) to be ideal for NFVI. Column generation allows us to place multiple VNF SCs in a relatively small amount of time; and to the best of our knowledge, it is the first to solve the multiple VNF SC placement and routing problem. 

\section{Problem Description}
\label{probDesc}

\begin{figure}
\center
 \includegraphics[scale=.35]{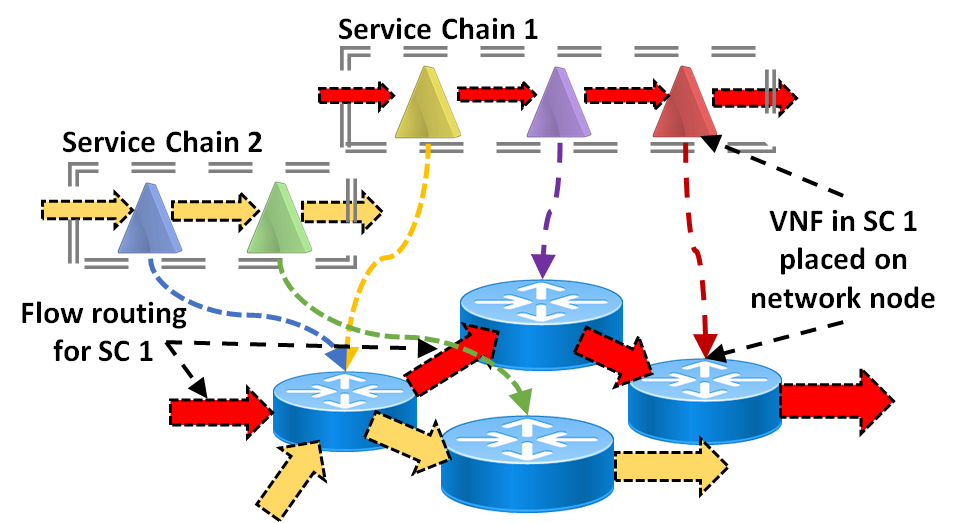}
 \caption{Multiple VNF SC placement and routing.}
  \label{fig:multiSC}
\end{figure}

An operator's network provides multiple services. Traffic demand for a service is satisfied by sequential traversal of VNFs comprising a SC for the service. Configuring VNFs to form an SC is an important problem but is not the focus of this work. Here, we assume that, given a service, the operator has a pre-configured SC, i.e., the ordered sequence of VNFs is already known. With the knowledge of SC configurations for each service, the network operator has to provide for the service requirements of all traffic flows with minimal network resources. The problem lends itself to be modeled as an optimization problem where the objective is to reduce network-resource consumption by optimally placing and routing traffic through multiple VNF SCs. Fig. \ref{fig:multiSC} depicts the multiple VNF SC placement and routing problem.

\subsection{Problem Statement}
\label{probState}

Given a network topology, capacity of links, a set of network nodes with NFV support (NFV-PoP), traffic flows for source-destination $(v_s,v_d)$ pairs, bandwidth demand of traffic flows for $(v_s,v_d)$ pairs, set of VNFs, and the set of SCs, we determine the placement of VNFs and corresponding routing to minimize network-resource (bandwidth) consumption.

\subsection{Input Parameters}
VNFs are distinct from VMs in that they are more bandwidth intensive. This happens as VNFs are virtual instances of network functions, which operate at line rate. Also, since a number of physical network functions are compute intensive, we assume VNF behaviour to be compute intensive. Each VNF has CPU core requirement to process unit traffic throughput. We assume that CPU core requirements of VNFs increase linearly with traffic throughput.

\begin{itemize}
\item $G = (V, L)$: Physical topology of the optical backbone network; $V$ is set of nodes and $L$ is set of links.
\item $V^{\NFV} \subseteq V$: Set of NFV-PoP nodes.
\item $F$, indexed by $f$: Set of VNFs.
\item $\ncore$: Number of CPU cores present per NFV-PoP node.
\item $\ncoref$: Number of CPU cores for function $f$ to give unit throughput.
\item $\ServiceChainSet$: Set of chains, indexed by $c$.
\item $n_c$: Number of VNFs in SC $\ServiceChainIdx$.
\item $\mathcal{SD}$: Set of source-destination $(v_s,v_d)$ pairs.
\item $D^c_{sd}$: Traffic demand between $v_s$ and $v_d$ for SC $\ServiceChainIdx$.
\item $\sigma_i(\ServiceChainIdx)$: ID of $i$th VNF in SC $\ServiceChainIdx$ where $f_{\sigma_i(\ServiceChainIdx)} \in F$.
\end{itemize}

As discussed in Section \ref{relWrk}, our previous model \cite{ilp_report} did not scale for multiple SCs and could not enforce the number of SC instances to be deployed. Hence, we have utilized a Decomposition Model (`Column Generation'), which scales for multiple VNF SC deployment and traffic routing.

\section{Decomposition Model}
\label{colGen}

Each SC, denoted by $\ServiceChainIdx$, is characterized by an ordered set of $n_c$ functions:
\begin{equation}
\text{[SC $\ServiceChainIdx$]} \qquad 
f_{\sigma_1(\ServiceChainIdx)} \prec f_{\sigma_2(\ServiceChainIdx)} \prec \dots \prec f_{\sigma_{n_\ServiceChainIdx}(\ServiceChainIdx)}.
\end{equation}
Each deployment of SC $\ServiceChainIdx$ is defined by a set of VNF locations, and a set of paths, from the location of first VNF to location of last VNF.

Our decomposition model relies on a set of \textit{chain configurations} where each configuration is associated with a potential provisioning of a SC $\ServiceChainIdx$ and a potential node placement of its functions. 
Let $\ConfigSet$ be the set of configurations, and $\ConfigSet_\ServiceChainIdx$ be the subset of configurations associated with service chain 
$\ServiceChainIdx \in \ServiceChainSet$:
$$\quad \ConfigSet = \bigcup\limits_{\ServiceChainIdx \in \ServiceChainSet} \ConfigSet_\ServiceChainIdx.$$ 
Due to its definition, the number of configurations is exponential. 
Given this consideration, we find the problem to fit naturally in the column-generation framework \cite{jaumard_colgen_rwa}. 

Column generation (\textbf{CG}) is a decomposition technique, where the problem (called Master Problem - \textbf{MP}) to be solved is divided into two sub-problems: restricted master problem (\textbf{RMP}) (selection of the best configurations) and pricing problems
 (\textbf{PP\_SC($ \ServiceChainIdx$)})$_{\ServiceChainIdx \in \ServiceChainSet}$ (configuration generators for each chain). 
The CG process involves solving the \textbf{RMP}, querying the dual values of \textbf{RMP} constraints, and using them for \textbf{PP\_SC($ \ServiceChainIdx$)} objective. Each improving solution (i.e., with a negative reduced cost) of \textbf{PP\_SC($ \ServiceChainIdx$)} is added to \textbf{RMP}, and previous step is repeated until optimality condition is reached \cite{jaumard_colgen_rwa}, with the \textbf{PP\_SC($ \ServiceChainIdx$)} explored in a round robin fashion.

A chain configuration is characterized by the following parameters: 
\begin{itemize}
\item Location of the functions: $a_{vf}^{\ConfigIdx} =1$ if $f \in \ServiceChainIdx$ is located in $v$ in configuration; 0 otherwise.
\item Connectivity of the locations: path from the location of current VNF to next VNF in SC $\ServiceChainIdx$. If link $\ell$ is used in the path from the location of $f_{\sigma_i(\ServiceChainIdx)}$ to the location of $f_{\sigma_{i+1}(\ServiceChainIdx)}$, then $b_{i \ell}^{\ConfigIdx} = 1$; 0 otherwise.
\end{itemize}

\subsection{Restricted Master Problem (\textbf{RMP}) }
\label{masterProb}

\textbf{RMP} selects the best $\ConfigIdx \in \ConfigSet_\ServiceChainIdx$ for each SC $\ServiceChainIdx$. Also it finds a route from $v_s$ (source) to first VNF of $\ServiceChainIdx$ and from last VNF of $\ServiceChainIdx$ to $v_d$ (destination).

\begin{figure}[htb]
\begin{center}
  \subfloat[][A first configuration ($\ConfigIdx_1$) for $c$]{\label{fig:Config1}\includegraphics[scale=.25]{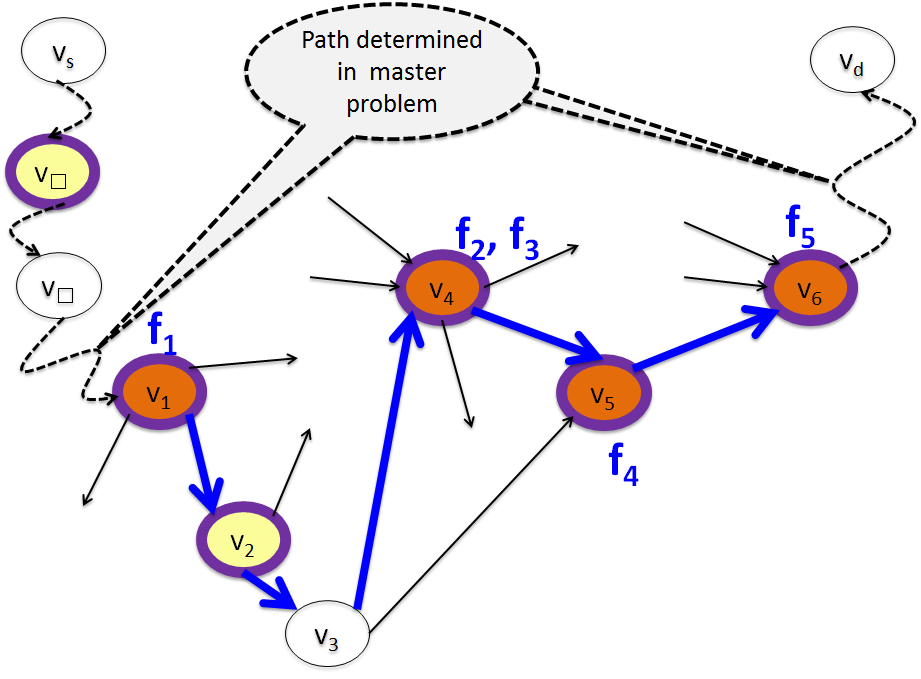}} \\
  \subfloat[][A second configuration ($\ConfigIdx_2$) for $c$]{\label{fig:Config2}\includegraphics[scale=.25]{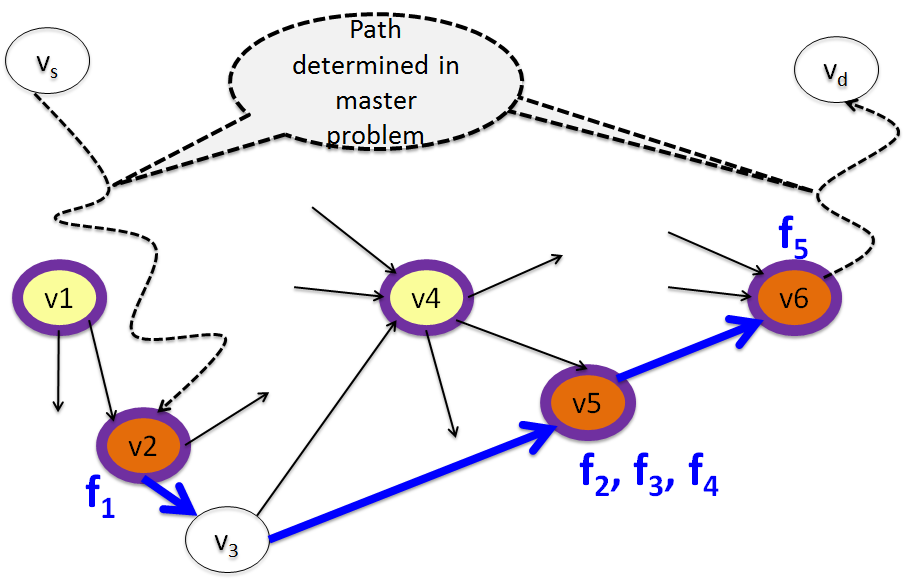}}
\end{center}
\caption{Two configuration examples for  chain $c = (f_1 \prec f_2 \prec f_3 \prec f_4 \prec f_5)$.}
\label{fig:service_chain}	
\end{figure}

An illustration of the constraint splitting between \textbf{RMP} and \textbf{PP\_SC($ \ServiceChainIdx$)} is depicted in Fig. \ref{fig:service_chain}. Nodes circled in purple are NFV-PoP, yellow nodes do not host VNFs at present but have NFV support, and orange nodes currently host VNFs. Figure \ref{fig:service_chain}\subref{fig:Config1} has $f_1$ located at $v_1$. When a different configuration is selected in Fig. \ref{fig:service_chain}\subref{fig:Config2} and $f_1$ is located at $v_2$, then \textbf{RMP} finds the path from $v_s$ to location of $f_1$. Similarly, \textbf{RMP} finds the path from last VNF to $v_d$, i.e., $f_5$ to $v_d$ here. 

\vspace{0.2cm}
\noindent
\textbf{Variables:}
\begin{itemize}
\item $z_{\ConfigIdx} =1$ if configuration $\ConfigIdx$ is selected; 0 otherwise.
\item $x_{v}^{f} =1$ if $f$ is located in $v$; 0 otherwise.
\item $y^{\text{first(\ServiceChainIdx)}, sd}_{\ell} =1$ if $\ell$ is on path from $v_s$ to location of first VNF in $\ServiceChainIdx$; 0 otherwise.
\item $y^{\text{last(\ServiceChainIdx)}, sd}_{\ell} =1$ if $\ell$ is on path from location of last VNF in $\ServiceChainIdx$ to $v_d$; 0 otherwise.
\end{itemize}

\vspace{0.2cm}
\noindent
\textbf{Objective:} Minimize bandwidth consumed:
\begin{multline} \min \quad 
     \sum\limits_{\ServiceChainIdx \in \ServiceChainSet} \>  \sum\limits_{\ConfigIdx \in \ConfigSet_{\ServiceChainIdx}}  \>
     \underbrace{ 
     \sum\limits_{\ell \in L} \> 
     \overbrace{ \left(   \sum\limits_{(s,d) \in \mathcal{SD}}   \> D^c_{sd}  \right) }^{\text{Overall traffic using } \ServiceChainIdx} 
                   \sum\limits_{i \in I} b_{i \ell}^{\ConfigIdx}
                   }_{\textsc{cost}_{\ConfigIdx}} 
                   \zRMP  + \\
     \sum\limits_{\ServiceChainIdx \in \ServiceChainSet} \> 
     \sum\limits_{\ell \in L} \> \sum\limits_{(s,d) \in \mathcal{SD}}   \> D^c_{sd}  \left( y^{\text{first(\ServiceChainIdx)}, sd}_{\ell} + y^{\text{last(\ServiceChainIdx)}, sd}_{\ell} \right).    
\end{multline}

Total bandwidth consumed in placing multiple SCs depends on configuration  $\ConfigIdx$  selected for each SC  $\ServiceChainIdx$. Each $\ConfigIdx$ for $\ServiceChainIdx$ locates VNFs of $\ServiceChainIdx$ and gives the route to traverse these VNF locations. So, bandwidth consumed when going from $v_s$ to $v_d$ and traversing the SC depends on selected $\ConfigIdx$.  

\vspace{0.2cm}
\noindent
\textbf{Constraints}:
\begin{alignat}{2}
 & \sum\limits_{{\ConfigIdx} \in \ConfigSet_\ServiceChainIdx} z_{\ConfigIdx} = 1  
 && \hspace*{-4.cm} \ServiceChainIdx \in \ServiceChainSet \label{eq:single_config_per_service_chain} \\
& \sum\limits_{\ConfigIdx \in \ConfigSet} \> \sum\limits_{v \in V^{\NFV}} a_{vf}^{\ConfigIdx} \>  \zRMP \geq 1
&& \hspace*{-4.cm} f \in F \label{eq:function_served} \\
& \sum\limits_{\ServiceChainIdx \in \ServiceChainSet} \> \sum\limits_{{\ConfigIdx} \in \ConfigSet_{\ServiceChainIdx}} \> \sum\limits_{f \in F} \ncoref  \sum\limits_{(v_s, v_d) \in \mathcal{SD}}  D_{sd}^c \> a_{vf}^{\ConfigIdx} \> \zRMP \leq \ncore 
&& \myspace \nonumber \\
& &&  \hspace*{-4.cm} v \in V^{\NFV} \label{eq:capa_cores} \\
&   \sum\limits_{\ServiceChainIdx \in \ServiceChainSet} \>  \sum\limits_{(v_s, v_d) \in \mathcal{SD}}  \> D^{\ServiceChainIdx}_{sd} \left( y^{\text{first(\ServiceChainIdx)}, sd}_{\ell} + y^{\text{last(\ServiceChainIdx)}, sd}_{\ell}     + \sum\limits_{{\ConfigIdx} \in \ConfigSet} \> \sum\limits_{i =1}^{n_{\ServiceChainIdx} - 1}  b_{i \ell}^{\ConfigIdx} \>    
   \zRMP \right)   && \nonumber \\
& \qquad \qquad \qquad \qquad 
  \leq \textsc{cap}_{\ell}  
&& \hspace*{-2.cm}  \ell \in L \label{eq:capacity} \\
& \sum\limits_{\ConfigIdx \in \ConfigSet} a_{vf}^{\ConfigIdx} \zRMP \geq x^f_{v}   
&&  \hspace*{-4.cm} f \in F, v \in V^{\NFV} \label{eq:a_x_consistent1} \\
& \sum\limits_{\ConfigIdx \in \ConfigSet} a_{vf}^{\ConfigIdx} \zRMP \leq M \> x^f_{v}  
&&  \hspace*{-4.cm} f \in F, v \in V^{\NFV}. \label{eq:a_x_consistent2} 
\end{alignat}  

Constraints \eqref{eq:single_config_per_service_chain} guarantee that we select exactly one $\ConfigIdx$ for SC  $\ServiceChainIdx$  
and forces $\ServiceChainIdx$ to have a single instance. 
Each $\ConfigIdx$ is associated with a set of $a^{\ConfigIdx}_{vf}$ (from \textbf{PP\_SC($ \ServiceChainIdx$)}) required to be consistent with $x_v^f$ in \textbf{RMP}, which is resolved by Eqs. \eqref{eq:a_x_consistent1} and \eqref{eq:a_x_consistent2}.

We also ensure that each VNF $f$ to be deployed across SCs is placed on at least one node using Eq. \eqref{eq:function_served}. While placing $f$ in a NFV-PoP node, we need to ensure that this node has sufficient CPU cores. This is done using Eq. \eqref{eq:capa_cores} with consideration for increase in compute resource demand due to traffic increase. 

Eq. \eqref{eq:capacity} enforces link-capacity constraints for the complete route for SC  $\ServiceChainIdx$  from $v_s$ to $v_d$ ( $\forall (v_s,v_d) \in \mathcal{SD} : D^{c}_{s,d} > 0$).    
\begin{alignat}{2}     
& \text{\textbf{Route from} } v_s \text{ \textbf{to first function location:}} \nonumber \\
& \sum\limits_{\ell \in \omega^+{(v_s)}} y^{\text{first(\ServiceChainIdx)}, sd}_{\ell} = 1 - x^{\text{first(\ServiceChainIdx)}}_{v_s}
&& \myspace  \ServiceChainIdx \in \ServiceChainSet, \nonumber \\
& &&  \hspace*{-3.5cm}  (v_s,v_d) \in \mathcal{SD}: D_{sd}^{\ServiceChainIdx} > 0 \label{eq2:link_from_source_to_ingress} \\
& \sum\limits_{\ell \in \omega^-{(v)}} y^{\text{first(\ServiceChainIdx)}, sd}_{\ell}  \geq x^{\text{first(\ServiceChainIdx)}}_v
&&  \hspace*{-2.cm}  \ServiceChainIdx \in \ServiceChainSet, \nonumber \\
& && \hspace*{-5.cm}  (v_s,v_d) \in \mathcal{SD}: D_{sd}^{\ServiceChainIdx} > 0, v \in V^{\NFV} \setminus \{ v_s \} \label{eq2:to_ensure_NFV_1_placement} \\
&     \sum\limits_{\ell \in \omega^+{(v)}} y^{\text{first(\ServiceChainIdx)}, sd}_{\ell}
     -  \sum\limits_{\ell \in \omega^-{(v)}} y^{\text{first(\ServiceChainIdx)}, sd}_{\ell} 
     = - x^{\text{first(\ServiceChainIdx)}}_v
&& \nonumber \\
& && \hspace*{-6.cm}  \ServiceChainIdx \in \ServiceChainSet, 
            (v_s,v_d) \in \mathcal{SD}: D_{sd}^{\ServiceChainIdx} > 0, v \in V^{\NFV} \setminus \{ v_s \}        \label{eq:places_NFV} \\
&     \sum\limits_{\ell \in \omega^+{(v)}} y^{\text{first(\ServiceChainIdx)}, sd}_{\ell}
     -  \sum\limits_{\ell \in \omega^-{(v)}} y^{\text{first(\ServiceChainIdx)}, sd}_{\ell}
     = 0
      && \nonumber \\
      & && \hspace*{-5.cm}  \ServiceChainIdx \in \ServiceChainSet, (v_s,v_d) \in \mathcal{SD}: D_{sd}^{\ServiceChainIdx} > 0, \nonumber \\
      & && \hspace*{-3.cm} v \in V \setminus (V^{\NFV} \cup \{ v_s \}).         \label{eq:places_non_NFV} 
\end{alignat} 

We assume that an unique route exists from $v_s$ to first VNF location. This is imposed by selecting exactly one outgoing link from $v_s$ unless first VNF is located at 
$v_s$. We account for these scenarios using Eq. \eqref{eq2:link_from_source_to_ingress}. To find the route from $v_s$ to first VNF, flow conservation needs to be enforced at the intermediate nodes which may or may not have NFV support. Eqs. \eqref{eq:places_NFV} and \eqref{eq:places_non_NFV} enforces flow-conservation constraints at nodes with and without NFV support, respectively.

Eq. \eqref{eq2:to_ensure_NFV_1_placement} ensures an incoming link to node hosting first VNF, unless that node is $v_s$.  
\begin{alignat}{2}
& \text{\textbf{Route from last function location to }} v_d \text{\textbf{:}}\nonumber \\
& \sum\limits_{\ell \in \omega^-{(v_d)}} y^{\text{last(\ServiceChainIdx)}, sd}_{\ell} 
    = 1 - x^{\text{last(\ServiceChainIdx)}}_{v_d}
&& \hspace*{-1.cm}  \ServiceChainIdx \in \ServiceChainSet, \nonumber \\
& && \hspace*{-3.cm} (v_s,v_d) \in \mathcal{SD}: D_{sd}^{\ServiceChainIdx} > 0  \label{eq2:link_from_egress_to_destination} \\      
& \sum\limits_{\ell \in \omega^+{(v)}} y^{\text{last(\ServiceChainIdx)}, sd}_{\ell} \geq 
     x^{\text{last(\ServiceChainIdx)}}_v  
&& \hspace*{-2.cm}  \ServiceChainIdx \in \ServiceChainSet, \nonumber \\
& && \hspace*{-5.cm} (v_s, v_d) \in \mathcal{SD}: D_{sd}^{\ServiceChainIdx} > 0, v \in V^{\NFV} \setminus \{ v_d \}  
      \label{eq2:to_ensure_NFV_N_placement} \\ 
&      \sum\limits_{\ell \in \omega^+{(v)}} y^{\text{last(\ServiceChainIdx)}, sd}_{\ell} 
     -   \sum\limits_{\ell \in \omega^-{(v)}} y^{\text{last(\ServiceChainIdx)}, sd}_{\ell}  = x^{\text{last(\ServiceChainIdx)}}_v && \nonumber \\
& && \hspace*{-6.cm}  \ServiceChainIdx \in \ServiceChainSet, (v_s, v_d) \in \mathcal{SD}: D_{sd}^{\ServiceChainIdx} > 0,  v \in V^{\NFV} \setminus \{ v_d \} \label{eq:places_NFV_destination} \\
&      \sum\limits_{\ell \in \omega^+{(v)}} y^{\text{last(\ServiceChainIdx)}, sd}_{\ell} 
     -   \sum\limits_{\ell \in \omega^-{(v)}} y^{\text{last(\ServiceChainIdx)}, sd}_{\ell}  = 0 && \nonumber \\
    & && \hspace*{-5.cm}  \ServiceChainIdx \in \ServiceChainSet, (v_s, v_d) \in \mathcal{SD}: D_{sd}^{\ServiceChainIdx} > 0,  \nonumber \\
    & && \hspace*{-3.cm} v \in V \setminus (V^{\NFV} \cup \{ v_d \}). \label{eq:places_not_NFV_destination}
\end{alignat}

Eqs. \eqref{eq2:link_from_egress_to_destination},  \eqref{eq:places_NFV_destination}, \eqref{eq:places_not_NFV_destination}, 
and \eqref{eq2:to_ensure_NFV_N_placement} enforce same functionality as Eqs. \eqref{eq2:link_from_source_to_ingress},  \eqref{eq:places_NFV}, \eqref{eq:places_non_NFV}, and \eqref{eq2:to_ensure_NFV_1_placement}, respectively, but on route from location of last VNF to $v_d$.

\subsection{Pricing Problem}
\label{pricingProb}

Each SC  $\ServiceChainIdx$  ($\ServiceChainIdx \in \ServiceChainSet$) is associated with a pricing problem. The number of pricing problems to be solved equals the number of SCs to be deployed. 

Pricing problem \textbf{PP\_SC($ \ServiceChainIdx$)} generates: 
\textit{(i)} A set of locations for VNFs of  $\ServiceChainIdx$; and
\textit{(ii)} a sequence of paths from the location of VNF $f_{\sigma_i( \ServiceChainIdx)}$ to the location of VNF $f_{\sigma_{i+1}( \ServiceChainIdx)}$, for $i = 1, 2, \dots, n_\ServiceChainIdx - 1$ for chain $\ServiceChainIdx$. Each solution that generated by \textbf{PP\_SC($\ServiceChainIdx$)} with a negative reduced cost, leads to a new potential $\ConfigIdx$ for  $\ServiceChainIdx$ of interest.

Let $u_{c}^{\eqref{eq:single_config_per_service_chain}} \lesseqgtr 0, u_f^{\eqref{eq:function_served}} \geq 0, u^{\eqref{eq:capa_cores}}_{v} \geq 0, u_{vf}^{\eqref{eq:a_x_consistent1}} \geq 0, u_{vf}^{\eqref{eq:a_x_consistent2}} \geq 0$ be the values of the dual variables associated with constraints \eqref{eq:single_config_per_service_chain}, \eqref{eq:function_served}, \eqref{eq:capa_cores}, \eqref{eq:a_x_consistent1}, and  \eqref{eq:a_x_consistent2}, respectively.

\vspace{.2cm} 

\noindent
\textbf{Variables}: 
\begin{itemize}
\item $a_{vf}$ = 1 if one occurrence of VNF $f$ is located in $v \in V^{\NFV}$; 0 otherwise. 
\item $ b_{i\ell}^{\ConfigIdx}$ = 1 if $\ell$ is on the path from location of  
$f_{\sigma_i(\ServiceChainIdx)}$ to location of $f_{\sigma_{i+1}(\ServiceChainIdx)}$; 0 otherwise.
\end{itemize}

\vspace{.2cm}

\noindent
\textbf{Objective:} Minimize reduced cost of variable $z_{\ConfigIdx}$: 
\begin{multline}
\text{[\textbf{PP\_SC($\ServiceChainIdx$)}]} \qquad 
 \redcost_{\ConfigIdx} = \cost_{\ConfigIdx}
-  \sum\limits_{f \in F}  u_f^{\eqref{eq:function_served}} \> \sum\limits_{v \in V} a_{vf} \\ 
+ \sum\limits_{v \in V^{\NFV}} \> u^{\eqref{eq:capa_cores}}_{v} \sum\limits_{f \in F_\ServiceChainIdx}  \ncoref \>{\sum\limits_{(v_s, v_d) \in \SD} }D_{sd}^{\ServiceChainIdx} \> a_{vf}  \\
+ \sum\limits_{\ell \in L} \sum\limits_{(v_s, v_d) \in \SD} u^{\eqref{eq:capacity}}_{\ell} D_{sd}^c 
                  \sum\limits_{i =1}^{n_{\ServiceChainIdx} - 1} b_{i\ell}^{\ConfigIdx} \\
- \sum\limits_{f \in F} \> \sum\limits_{v \in V^{\NFV}} u_{vf}^{\eqref{eq:a_x_consistent1}} a_{vf}       
+ \sum\limits_{f \in F} \> \sum\limits_{v \in V^{\NFV}} u_{vf}^{\eqref{eq:a_x_consistent2}} a_{vf} 
- u_{c}^{\eqref{eq:single_config_per_service_chain}}.                
\end{multline}

\noindent
where $\redcost$ value indicates whether an optimal $\ConfigIdx$ for $\ServiceChainIdx$ has been found. A non-negative value of $\redcost$ indicates optimality for our model.

\vspace{.1cm}
\noindent 
\textbf{Constraints:}
\begin{alignat}{2}
& \sum\limits_{f \in F_{\ServiceChainIdx}} n_f^{\core} \sum\limits_{(v_s, v_d) \in \SD} D^{\ServiceChainIdx}_{sd} \> a_{vf} \leq \ncore
&& \myspace   v \in V^{\NFV} \label{eq:PP_node_capacity} \\
& \sum\limits_{(v_s,v_d) \in \SD} D^{c}_{sd} \sum\limits_{i=1}^{n_c - 1} b_{\ell}^{\sigma_{i}(c), {\sigma_{i+1}(c)}} \leq \CAP_{\ell}    
&& \myspace    \ell \in L \label{eq:PP_link_capacity} \\
& \sum\limits_{v \in V^{\NFV}} a_{v \sigma_i(\ServiceChainIdx)} = 1
&& \myspace \hspace*{-1.cm}  i = 1, 2, \dots, n_{\ServiceChainIdx}  \label{eq:each_function} \\
& \sum\limits_{\ell \in \omega^+ (v)} b_{1 \ell}^{\ConfigIdx} \geq a_{v, \sigma_1(\ServiceChainIdx)} 
- a_{v, \sigma_2(\ServiceChainIdx)}
&& \myspace v \in V^{\NFV} \label{eq:outgoing1} \\
& \sum\limits_{\ell \in \omega^+ (v)}  b_{1 \ell}^{\ConfigIdx} \leq 1 - a_{v, \sigma_2(\ServiceChainIdx)} 
&& \myspace v \in V^{\NFV} \label{eq:incoming2}\\
& \sum\limits_{\ell \in \omega^- (v)}  b_{1 \ell}^{\ConfigIdx} \leq 1 - a_{v, \sigma_1(\ServiceChainIdx)} 
&& \myspace v \in V^{\NFV} \label{eq:incoming1}\\
&    \sum\limits_{\ell \in \omega^+(v)} b_{i \ell}^{\ConfigIdx} 
    - \sum\limits_{\ell \in \omega^-(v)}  b_{i \ell}^{\ConfigIdx} 
      = a_{v, \sigma_i(\ServiceChainIdx)} - a_{v, \sigma_{i+1}(\ServiceChainIdx)} \nonumber \\
& && \hspace*{-5.cm} v \in V^{\NFV}, i = 1, 2, \dots, n_{\ServiceChainIdx} - 1 \label{eq:flowNFV}\\
& \sum\limits_{\ell \in \omega^+(v)} b_{i \ell}^{\ConfigIdx} - \sum\limits_{\ell \in \omega^-(v)} b_{i \ell}^{\ConfigIdx} = 0 && \nonumber \\
& && \hspace*{-5.cm} v \in V \setminus V^{\NFV},  i = 1, 2, \dots, n_{\ServiceChainIdx} - 1 \label{eq:flow} \\
&   \sum\limits_{\ell \in \omega^- (v)} 
    b_{(n_{\ServiceChainIdx}-1) \ell}^{\ConfigIdx}  \geq 
    a_{v, \sigma_{n_{\ServiceChainIdx}}(\ServiceChainIdx)} - a_{v, \sigma_{n_{\ServiceChainIdx}-1}(\ServiceChainIdx)}  && \nonumber \\
&& \myspace v \in V^{\NFV} \label{eq:incominglast}\\
& \sum\limits_{\ell \in \omega^+ (v)}b_{(n_{\ServiceChainIdx}-1) \ell}^{\ConfigIdx} \leq 1 - a_{v, \sigma_{n_{\ServiceChainIdx}}(\ServiceChainIdx)} 
&& \myspace v \in V^{\NFV}. \label{eq:outgoinglast}
\end{alignat} 

Eqs. \eqref{eq:PP_node_capacity} and \eqref{eq:PP_link_capacity} are compute resource and capacity constraints, similar to those in \textbf{RMP}.

Eq. \eqref{eq:each_function} ensures that each VNF of SC  $\ServiceChainIdx$  is placed at least once. Eq. \eqref{eq:outgoing1} ensures that, if $f_{\sigma_1}(\ServiceChainIdx)$ is located in $v$, then a flow $b$ is outgoing from $v$ (one flow out, but no incoming flow in $v$). In addition, if $f_{\sigma_1}(\ServiceChainIdx)$ is not located in $v$, then Eq. \eqref{eq:outgoing1} is redundant. On the other hand, Eq. \eqref{eq:incoming1} ensures that, if $f_{\sigma_1}(\ServiceChainIdx)$ is located in $v$, then there is no flow ($b$) that is incoming to $v$. Also, if $f_{\sigma_1}(\ServiceChainIdx)$ is not located in $v$, then Eq. \eqref{eq:incoming1} is redundant.

Eqs. \eqref{eq:incominglast} and \eqref{eq:outgoinglast} are similar to Eqs. \eqref{eq:incoming1} and \eqref{eq:outgoing1}, but are related to last VNF in SC  $\ServiceChainIdx$.

Eqs. \eqref{eq:flowNFV} and \eqref{eq:flow} are flow-conservation constraints: Eq. \eqref{eq:flowNFV} for nodes with NFV support and Eq. \eqref{eq:flow} for other nodes.

\subsection{Solution Scheme}
The \textbf{PP\_SC($\ServiceChainIdx$)} are solved in a round-robin fashion and the final \textbf{RMP} is solved as an ILP, as shown in \cite{jaumard_colgen_rwa}.

\section{Illustrative Numerical Examples}
\label{sim_examples}

\begin{figure}[ht]
\center
 \includegraphics[scale=.2]{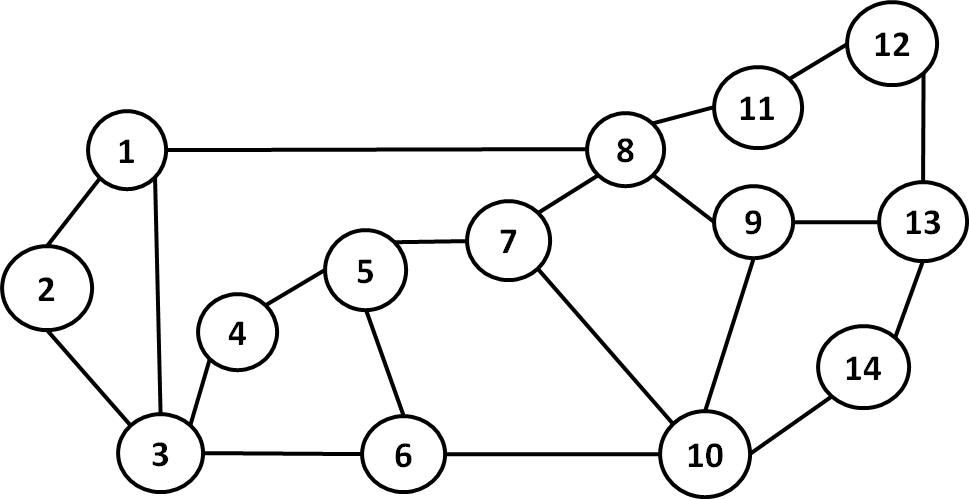}
 \caption{NSFNet topology.}
  \label{fig:topology}
\end{figure}

\begin{figure*}
  \centering
  \begin{tabular}{cc}
     \subfloat[][]{\label{fig:a}\includegraphics[width=.49\textwidth, scale=1]{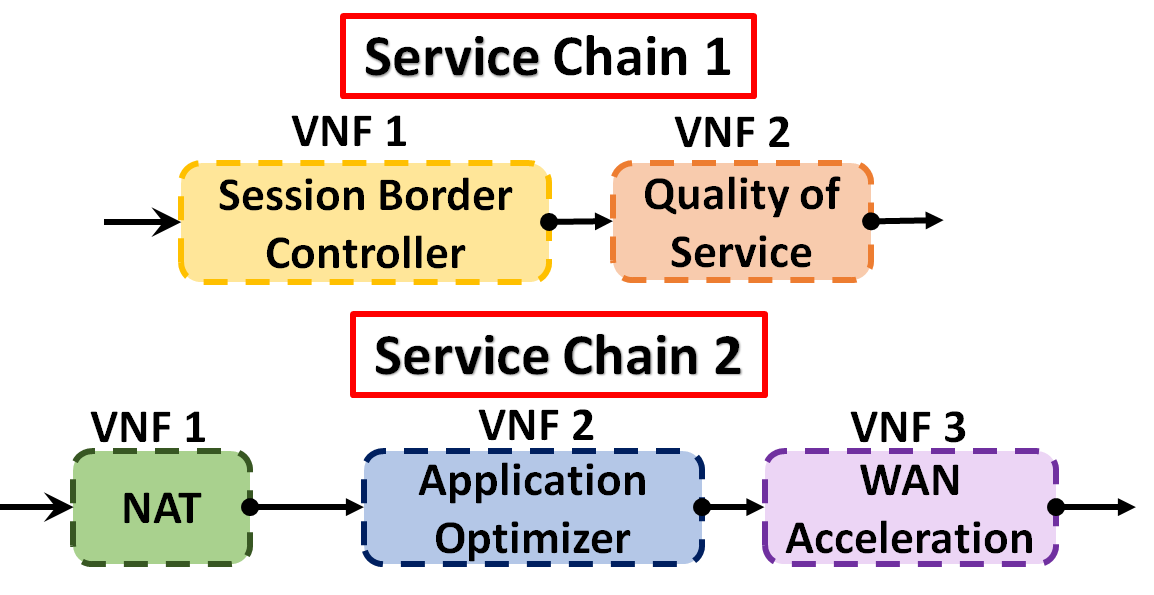}}
    & \subfloat[][]{\label{fig:b}\includegraphics[width=.49\textwidth, scale=1]{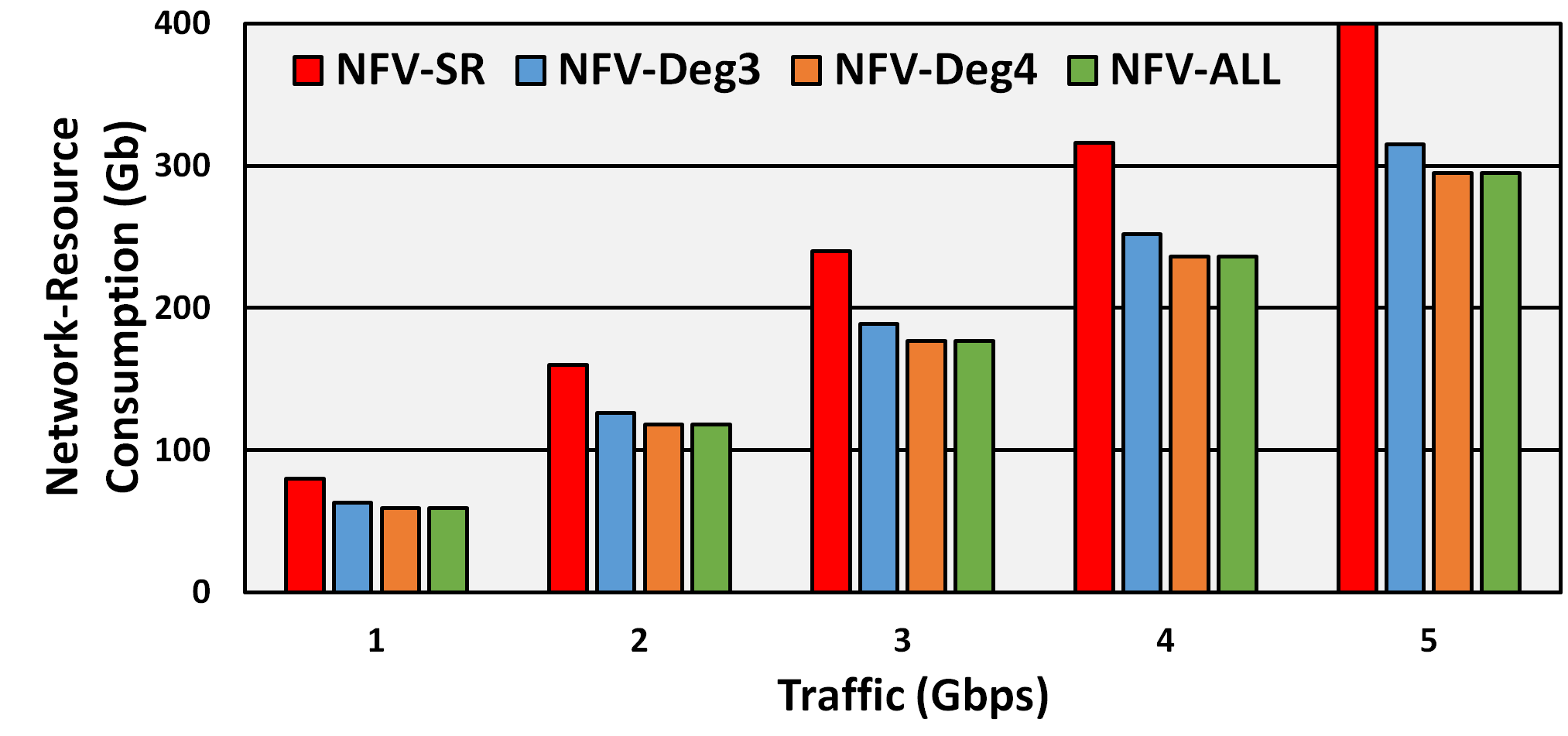}}
    \\ \\
     \subfloat[][]{\label{fig:c}\includegraphics[width=.49\textwidth, scale=1]{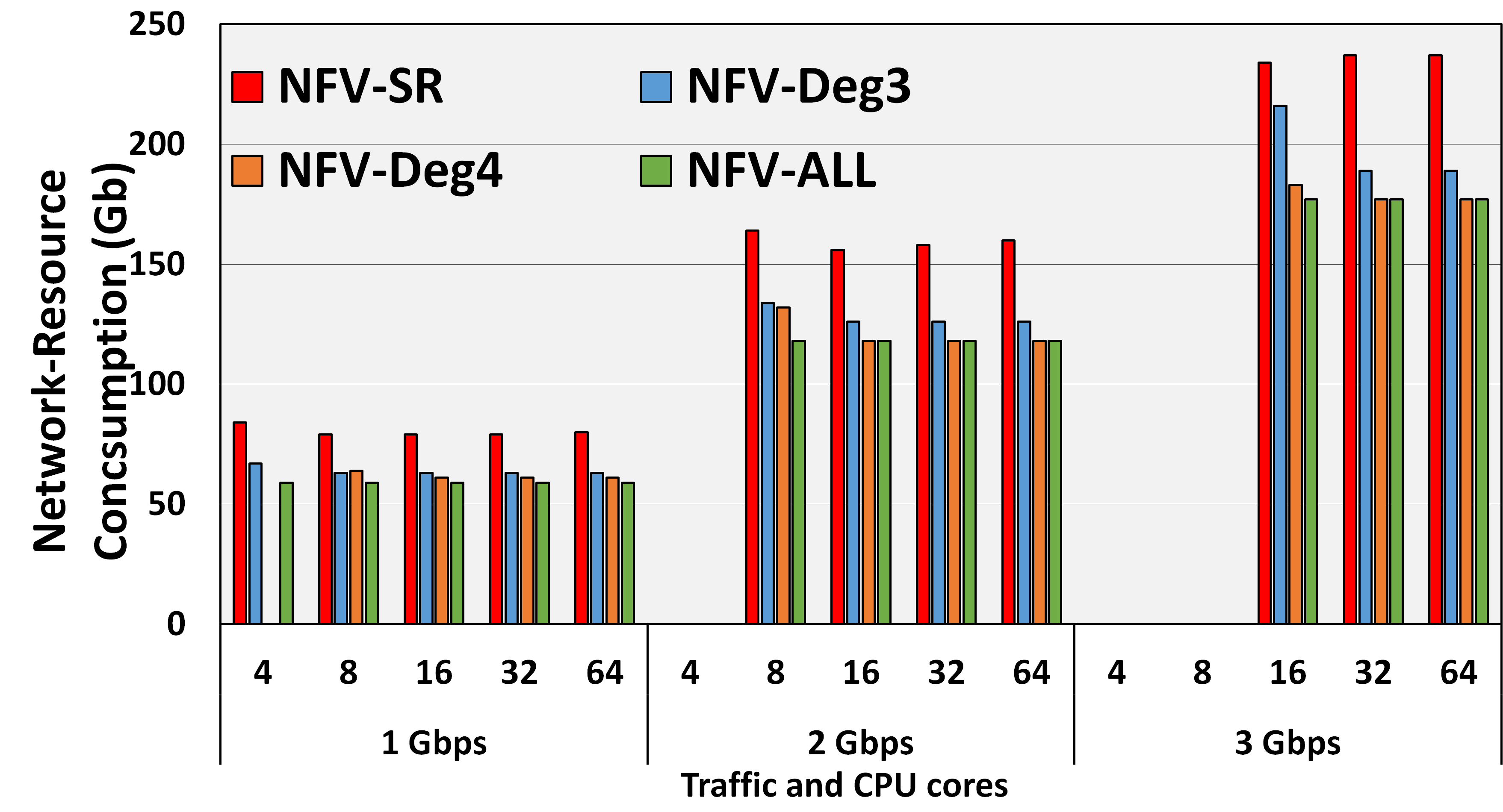}}
    & \subfloat[][]{\label{fig:d}\includegraphics[width=.49\textwidth, scale=1]{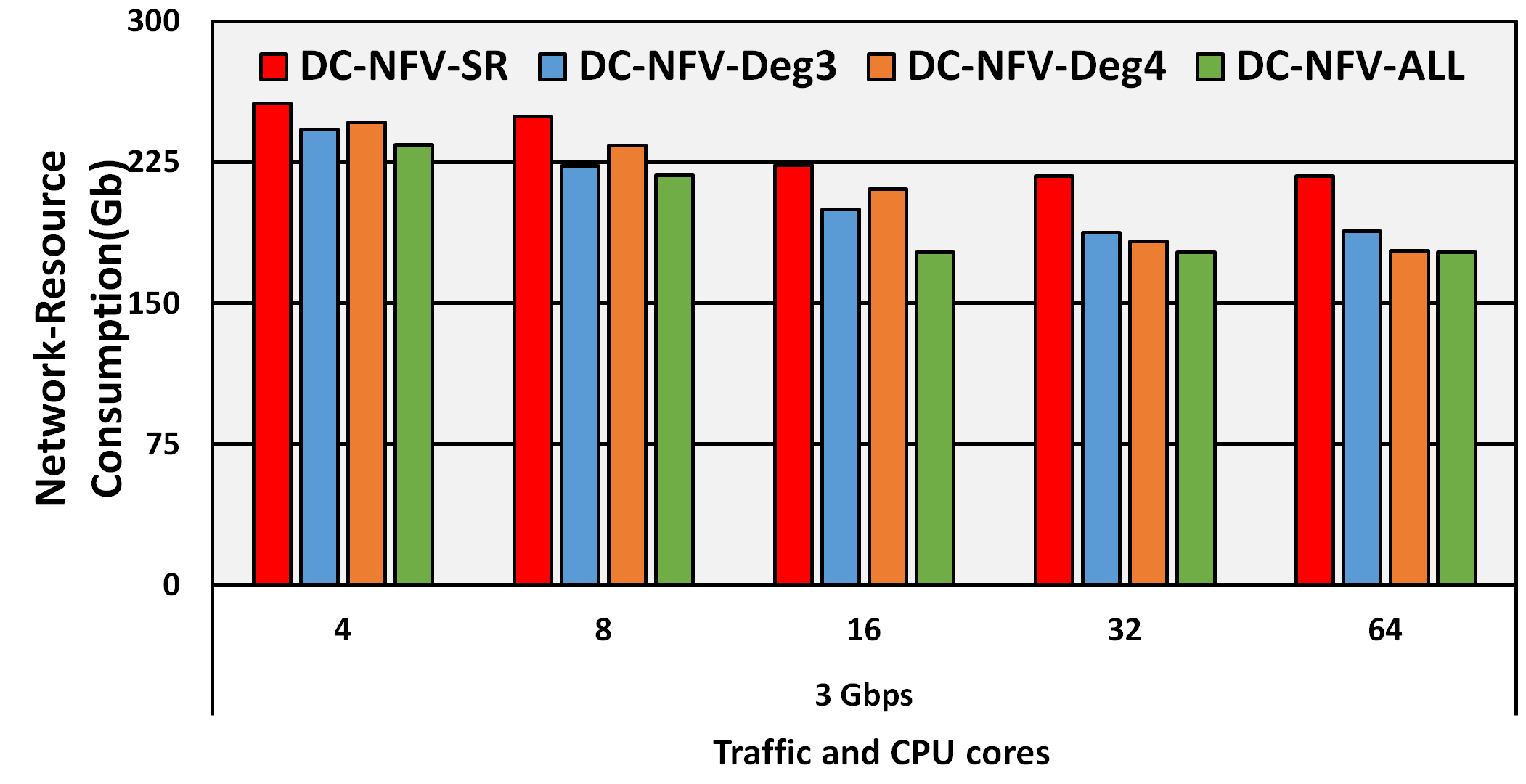}}
  \end{tabular}
  \caption{Example of service chains deployed and results.}
  \label{results}
\end{figure*}

We run instances of our optimization model over the NSFNet topology shown in Fig. \ref{fig:topology}. Each network link has 40 Gbps capacity in each direction. Fig. \ref{results}\subref{fig:a} depicts one of the type of SCs deployed for Figs.   \ref{results}\subref{fig:b}, \ref{results}\subref{fig:c} and \ref{results}\subref{fig:d}. Service Chain 1 is made up of Session Border Controller (SBC) and Quality of Service (QoS) VNFs which need to be traversed in sequence. Together, these two VNFs provide VoIP service with call-quality monitoring. We deploy 13 different service chains (SCs) over 20 uniformly-distributed traffic flows. All SCs comprise between two to five VNFs. In total, 33 different VNFs are used across the 13 SCs. This is a large number of distinct SCs and VNFs to be deployed. The solution for these problem instances is found in a relatively short time. The VNFs differ in functionality and CPU core requirements as given in \cite{cisco_vnf}.

Here, we employ the following five NFV infrastructure deployment schemes and compare their network resource consumption across varying traffic and CPU cores. 
\begin{itemize}
\item \textbf{NFV-Deg3}: In this scheme, all the degree-3 nodes in the network are made NFV-PoP. For our topology, \textbf{NFV-Deg3} will have nodes 1, 5, 6, 7, 9, and 13 as NFV-PoPs, i.e., they have NFV support.  
\item \textbf{NFV-Deg4}: All degree-4 nodes in the network are made NFV-PoP. In NSFNET, \textbf{NFV-Deg4} will have nodes 3, 8, and 10 as NFV-PoPs.
\item \textbf{NFV-ALL}: Here, all nodes in NSFNet are NFV-PoPs.
\item \textbf{NFV-SR}: Here, we are \textit{skewed} (S) towards nodes in the \textit{right} (R) of the network. So, nodes 9, 11, 12, 13, and 14 are made NFV-PoP. This scheme demonstrates how location of nodes is crucial for NFV-PoPs.
\item \textbf{DC-X}: Here `X' is any of the above schemes. This scheme represents a Network-enabled Cloud (NeC), as described in Section \ref{intro}. Here, a single DC node having massive compute resources is placed in the network and NFV-PoPs nodes are selected based on scheme `X'. To emulate a DC, we do not have CPU core capacity constraints at DC node. In our study, we place the DC node at each of the 14 nodes of NSFNet across our model instances. After each DC placement, NFV-PoPs are selected based on `X'. The final result is the average of results obtained from each DC placement.
\end{itemize}  

As number of nodes with NFV support in the network increases, traffic flows can satisfy service demands closer to the shortest path. However, the location of NFV-PoPs is as crucial. We demonstrate this with Fig. \ref{results}\subref{fig:b} where \textbf{NFV-SR} and \textbf{NFV-Deg3} have more NFV-PoPs (5 and 6, respectively) than \textbf{NFV-Deg4} (3 NFV-PoPs) but still have higher network-resource consumption. This happens as central nodes (as in \textbf{NFV-Deg4}) have more traffic flows pass through them. Also, \textbf{NFV-ALL} has least network-resource consumption since all nodes in the network in this scheme are NFV-PoP. Note that we have not enforced CPU core capacity constraints here.

A NFV-PoP has limited compute resources, and VNF placement will be contingent on resources available. Fig. \ref{results}\subref{fig:c} shows NFV-PoPs with different CPU cores. We find that, as traffic increases, the network-resource consumption trend is the same as in Fig. \ref{results}\subref{fig:b} with \textbf{NFV-SR} using the most network resources while \textbf{NFV-ALL} uses the least. However, we find that infeasibility occurs for schemes at certain traffic and CPU core ($\ncore$) values. \textbf{NFV-Deg4} becomes infeasible at $\ncore=4$ at 1 Gbps,  while all schemes are infeasible for $\ncore=4$ at 2 Gbps and  $\ncore=4,8$ at 3 Gbps. This happens due to increasing CPU core demand as traffic increases. If the NFV-PoP's CPU cores are not sufficient, the scenario is not feasible. So, 2 Gbps traffic becomes infeasible at $\ncore=4$ while 1 Gbps traffic does not. Infeasibility becomes a significant challenge with increasing traffic. 

 A massive compute resource node (e.g., a Data Center (DC)) in the network can avoid infeasibility as traffic can be routed to DC when NFV-PoPs have insufficient compute resources. Fig. \ref{results}\subref{fig:d} shows that \textbf{DC-X} scheme's  make infeasible $\ncore$ values of Fig. \ref{results}\subref{fig:c} become feasible. At $\ncore=4$ and 3 Gbps, all NFV-PoP schemes were infeasible in Fig. \ref{results}\subref{fig:c}. When we add a DC node to each scheme, we find that they become feasible for $\ncore=4$ and 3 Gbps in Fig. \ref{results}\subref{fig:d}. Similar observation is made for $\ncore=8$. This becomes possible as the traffic can be redirected to DC when compute resources are scarce.   

Our discussion on Figs. \ref{results}\subref{fig:b}, \ref{results}\subref{fig:c}, and \ref{results}\subref{fig:d} makes the case for a Network-enabled Cloud (NeC) for Network Function Virtualization Infrastructure (NFVI). A NeC will avoid resource constraints encountered in a NFVI having NFV-PoPs exclusively. Also, NeC can avoid congestion arising from a centralized NFVI (like a DC) since NFV-PoPs can share traffic load which makes NeC the ideal NFV Infrastructure (NFVI).

\section{Conclusion}
\label{concl}
We introduced the problem of service-chain placement and how, with Network Function Virtualization (NFV), we can use Virtual Network Functions (VNFs) for more flexible/agile service chaining. We developed a decomposition model for placement of multiple VNF service chains (SCs) and routing of traffic flows. Our model helps in solving this complex problem in a relatively small amount of time. Our objective was to demonstrate the reduction in network-resource consumption if we have more NFV Points of Presence (NFV-PoPs). We further showed that a Network-enabled Cloud (NeC) would be ideal for NFV Infrastructure (NFVI) since it avoids the resource constraints of a NFVI with NFV-PoPs only. NeC also avoids congestion problem of centralized NFVI (e.g., Data Center) since it has NFV-PoPs to share the service traffic load.

\section*{Acknowledgment}
This work was supported by NSF Grant No. CNS-1217978.



%

\bibliographystyle{atiq6}
\bibliography{gupta_nfv}

\begin{thebibliography}{10}

\bibitem{etsi_nfv}
ETSI, ``Network functions virtualisation: Introductory white paper.''
  \url{portal.etsi.org/NFV/NFV_White_Paper.pdf}, 2012.

\bibitem{rt_cloud}
Ericsson, ``The real-time cloud - combining cloud, {NFV} and service provider
  {SDN}.'' \url{www.ericsson.com/res/docs/whitepapers/wp-sdn-and-cloud.pdf},
  2014.

\bibitem{ietf_sc}
IETF, ``Network service chaining problem statement.''
  \url{https://tools.ietf.org/html/draft-quinn-nsc-problem-statement-00}, 2013.

\bibitem{vnf_placement_turck}
H. Moens and F. De~Turck, ``{VNF-P: A model for efficient placement of
  virtualized network functions},'' in \emph{Proc.} {\em 10th International
  Conference on Network and Service Management (CNSM)}, pp.~418--423, Nov.
  2014.

\bibitem{place_vnf_secci}
B. Addis, D. Belabed, M. Bouet, and S. Secci, ``Virtual network functions
  placement and routing optimization,'' {\em
  https://hal.inria.fr/hal-01170042/}, 2015.

\bibitem{vnf_placement_barcellos_gaspary}
M.~C. Luizelli, L.~R. Bays, L.~S. Buriol, M.~P. Barcellos, and L.~P. Gaspary,
  ``{Piecing together the NFV provisioning puzzle: Efficient placement and
  chaining of virtual network functions},'' in \emph{Proc.} {\em IFIP/IEEE
  International Symposium on Integrated Network Management (IM)}, pp.~98--106,
  May 2015.

\bibitem{orc_vnf_boutaba}
M. Bari, S. Chowdhury, R. Ahmed, and R. Boutaba, ``On orchestrating virtual
  network functions in {NFV},'' {\em Computing Research Repository},
  vol.~abs/1503.06377, 2015.

\bibitem{ilp_report}
A. Gupta, M. Habib, P. Chowdhury, M. Tornatore, and B. Mukherjee, ``{Joint
  Virtual Network Function Placement and Routing of Traffic in Operator
  Networks},'' {\em Technical Report, UC Davis}, 2015.

\bibitem{jaumard_colgen_rwa}
B. Jaumard, C. Meyer, and B. Thiongane, ``On column generation formulations for
  the {RWA} problem,'' {\em Discrete Applied Mathematics}, vol.~157,
  pp.~1291--1308, 2009.

\bibitem{cisco_vnf}
Cisco, ``{Cisco Cloud Services Router 1000V 3.14 Series Data Sheet}.''
  \url{http://www.cisco.com/c/en/us/products/collateral/routers/cloud-services-router-1000v-series/datasheet-c78-733443.pdf},
  2015.

\end{thebibliography}

\end{document}